\documentclass[3p,twocolumn,11pt]{elsarticle}
\usepackage{graphicx}
\usepackage{amssymb}
\usepackage{amsmath}
\usepackage{amsfonts}
\usepackage{color,url}
\usepackage[colorlinks=true,urlcolor=blue,anchorcolor=blue
,citecolor=blue,filecolor=blue,linkcolor=blue,menucolor=blue
,linktocpage=true,pdfproducer=medialab,pdfa=true]{hyperref}

\RequirePackage{xspace}

\newcommand{\MADGRAPH} {\textsc{MadGraph}\xspace}

\newcommand{\PYTHIA} {\textsc{pythia}\xspace}

\evensidemargin \oddsidemargin
\journal{Physics Letters B}

\begin{document}

\begin{frontmatter}

\title{An alternative method for searching for dimuon displaced vertices in the short-range region with ATLAS and CMS}
\author{A. Santocchia}
\affiliation[a]{INFN Sezione di Perugia and Perugia University, Italy}
\address{Via Pascoli, 06123-Perugia, Italy}
\begin{abstract}
Search for BSM phenomena is one of the fundamental goals of the LHC experiments. Many BSM models
foresee long-lived particles which decay far from the production vertex and a big effort
has been done both by the ATLAS and CMS collaborations to detect these long-lived particles.
A specific, widely studied, example for such searches is made through the measurement
of displaced vertices produced by a narrow resonance decaying to a muon pair. This paper
will first analyze what has been done so far and will try to evaluate if there are improvements
which may allow enhancing the discovery potential at LHC for such particles. 
The outcome of this work is that while long-lived particles decaying more than 1 cm from the 
primary vertex have been carefully studied by ATLAS and CMS, the region within a few millimeters 
from the primary vertex may have been partially neglected. 
A new approach for searching such resonances, fully based on data, will then be proposed which allows minimizing 
systematic uncertainties while keeping at a maximum the discovery potentiality for long-lived neutral
resonances which decay a few millimeters from the primary vertex.

\end{abstract}


\end{frontmatter}


\section{Introduction}
\label{sec:intro}

The standard model (SM) of particle physics is one of the most powerful theory 
ever developed by mankind. His success has been experimentally confirmed in the past 30 years 
in many sectors, the last one being the discovery of the Higgs boson at the LHC~\cite{ATLAS:2012yve,CMS:2012qbp}. 
There are however many aspects that suggest that the SM is not the end of the story and a 
tremendous effort has been made to find new phenomena beyond the SM which could give a hint 
on how to extend it.
One of the sectors where a big effort has been made 
recently by ATLAS~\cite{ATLAS:2008xda}, CMS~\cite{CMS:2008xjf} and LHCb~\cite{LHCb:2008vvz}, 
is the search of long-lived particles (LLPs). A LLP is an unstable particle with a 
decay length big enough to decay far from the production vertex and which travels 
a length measurable by a detector. There are many models (see~\cite{Alimena:2019zri} for a comprehensive review
on theory and experiment results on LLP)
which foresee such particles and the interest in such searches is very high.

Neutral LLPs can produce unique signatures, 
such as displaced vertices (DV) with respect to the primary vertex (PV), 
which could be detected by looking at the decay products. 
A simple option is to consider 
LLPs producing a narrow resonance decaying to $\mu^+\mu^-$. Generally speaking, this choice 
doesn’t limit the power of the discussed method because it could be applied also to other 
decay channels where a displaced vertex could be detected. 

In this paper I will first discuss what has been done up to now for detecting LLPs at 
ATLAS and CMS analyzing the different techniques used by both experiments. Subsequently, 
a reference model implemented in \PYTHIA ~\cite{Sjostrand:2006za} will be introduced and discussed, 
highlighting why it can be used to generate a LLP decaying to a lepton pair and why all the results 
obtained with this process could be used also for other models. Detector effects will 
be simulated using the Delphes~\cite{deFavereau:2013fsa} parametric simulation with a 
dedicated configuration card used to emulate the performance of the CMS detector. Finally, 
simulated events will be used to show that there are LLPs decaying to $\mu^+\mu^-$ which could go 
undetected with the methods used to date at LHC experiments 
and a new technique will be suggested to cover this unexplored area.


\section{Review of LLP searches at LHC}
\label{sec:review}
There are two different complementary methods for searching LLP: inclusive or exclusive searches.
In the first case, the search is mainly based on looking for a displaced vertex reconstructed 
from a muon pair or from one or more jets using their charged 
components measured in the tracking detector. 
The total invariant mass, built from the charged 
particles associated to the vertex, or the number of events detected are then used to look 
for any excess with respect to SM predictions. 
For exclusive searches, more complex topologies are tested, 
and the DV is associated with other leptons, jets or missing transverse energy. 
Topologies where the DV is not directly detectable because the decay products are neutral are also studied. 
Most of the time, however, inclusive and exclusive searches look for a DV:
the farther the DV is from the PV, 
the more difficult is to measure it because of the deterioration of the tracking efficiency. 
On the other hand, there are very few, well known, SM particles which produce a DV within 
LHC detectors acceptance so all these searches are almost background free.

\begin{figure}[hbtp]
  \begin{minipage}{0.45\textwidth}  
    \centering
    \includegraphics[width=0.98\textwidth]{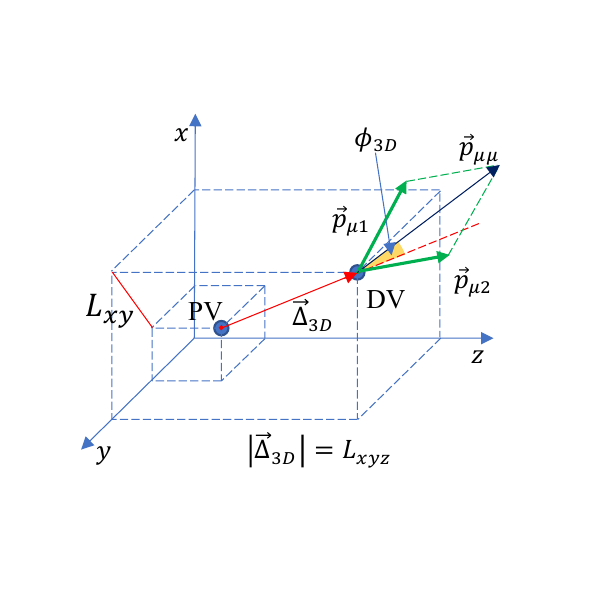}
    \caption{Schematic view of the $\phi_{3D}$ angle, $L_{xy}$ and $L_{xyx}$.}
    \label{fig:PV_DV}
  \end{minipage}
\end{figure}

This paper discusses only the inclusive search of a neutral, narrow resonance 
decaying to a muon pair. Both ATLAS and CMS use a set of standard requirements 
to guarantee the quality of the muon sample.
Special requirements are also used to minimize the 
contribution from cosmic ray muons which may be reconstructed as back-to-back muons. 
In the case of the two muons inclusive LLP searches, the most common additional requirements 
are on the muon transverse impact parameter $d_0$ or its significance $|d_0/\sigma_d|$ where $\sigma_d$ 
is the error on $d_0$.

After the selection of the muon pair, a secondary vertex finding algorithm may be run. 
If the vertex is found, then typical additional requirements are on 
the distance $L_{xy}$ in the transverse plane of the DV with respect to the 
PV or its significance $ L_{xy}/\sigma_{L_{xy}} $ where $\sigma_{L_{xy}}$ 
is the error on $ L_{xy}$. The distance $L_{xyz}$ of the DV with respect to the PV 
or its significance $L_{xyz}/\sigma_{L_{xyz}}$ where $\sigma_{L_{xyz}}$ is the error on $L_{xyz}$ is not used in 
any of the papers considered in this review. The main reason is probably the poor resolution of 
the longitudinal coordinate of the DV with respect to the resolution of the distance in the transverse plane. 
Another aspect that could drive this choice is the presence of pile-up events which could reduce the
efficiency of selecting the correct PV: the PV-DV distance in the transverse plane does not depend
on the presence of additional PVs along the beam direction while the 3D distance would change
dramatically if the wrong PV is selected. 
Another kinematic quantity used is $\phi_{3D}$: the angle between the 
PV-DV vector $\vec{\Delta}_{3D}$ (which represents the line of flight of the particle) 
and the momentum direction of the muon pair $\vec{p}_{\mu\mu}$. In the hypothesis of a LLP 2-body decay, this 
angle would be zero and any deviation would depend on the tracking resolution of the detector only.
Figure~\ref{fig:PV_DV} shows a schematic view of these quantities and
table~\ref{tab:inclusiveSearch} reports a summary of the main pre-selection cuts applied in 
all the papers published until 2023 by ATLAS and CMS for run 1 and run 2.

\begin{table*}[t]
  \centering
  \begin{tabular}{|l|c|c|c|c|c|c|c|c|c|c|}
  \hline
  Paper & Year & $\sqrt{s}$ & $\mathcal{L}$ fb$^{-1}$ & $|d_0|$& $|d_0|/\sigma_{d_0}$ 
        & $\chi^2_{ \mathrm{vtx}} $ fit & $L_{xy}$ & $L_{xy}/\sigma_{L_{xy}}$ & $\phi_{3D}$ \\
  \hline 
  CMS 1~\cite{CMS:2012axw} & 2013 & 7 & 5.1 & - & 2 & 5 & - & 5 & NO \\
  \hline
  CMS 2~\cite{CMS:2014hka} & 2015 & 8  & 20.5 & - & 12 & 5 & - & - & 2D \\
  \hline
  ATLAS 1~\cite{ATLAS:2018rjc} & 2019 & 13 & 32.9 & - & - & NO & - & - & NO \\
  \hline
  ATLAS 2~\cite{ATLAS:2019fwx} & 2020 & 13 & 32.8 & 2 & - & 5 & 2 & - & NO \\
  \hline
  ATLAS 3~\cite{ATLAS:2020xyo} & 2020 & 13 & 136 & 2 & - & 5 & 4 & - & NO \\
  \hline
  CMS 3~\cite{CMS:2021kdm} & 2022 & 13 & 113 & 0.1 & - & NO & - & - & NO \\
  \hline
  CMS 4~\cite{CMS:2021sch} & 2022 & 13 & 101 & - & 2/1 & 5 & - & - & 2D\\ 
  \hline
  CMS 5~\cite{CMS:2022qej} & 2023 & 13 & 97.6 & - & 6 & 10 & - & 6 & 2D \\
  \hline
  ATLAS 4~\cite{ATLAS:2023ios} & 2023 & 13 & 139 & 0.1 & - & NO & - & - & NO \\
  \hline
  \end{tabular}
  \caption{\label{tab:inclusiveSearch} CMS and ATLAS muon inclusive searches 
  ($d_0$, $\sigma_{d_0}$, $L_{xy}$ and $\sigma_{L_{xy}}$ are measured in mm). 
  2D in the table indicates that the $\phi_{3D}$ angle is used in the transverse plane.}
\end{table*}


\section{Analysis Strategy}
The capability to detect a BSM particle thanks to a DV depends on particle properties 
and detector performances. Specifically, the number of events with a DV 
that a detector could measure may be calculated through
\begin{equation}
  \label{eq:numberDV}
  N=\mathcal{L}\sigma \mathit{B}_{\mu\mu} \epsilon_D,
\end{equation}
where $N$ is the number of events detected, $\mathcal{L}$ is the integrated luminosity, $\sigma$ is the 
production cross-section of the resonance, $\mathit{B}_{\mu\mu} $ is the branching fraction 
of the resonance decaying to $\mu^+\mu^-$, and $\epsilon_D$ is the detector efficiency 
(including the efficiency of the vertex finding algorithm) for the specific process.
Going to a more detailed analysis, it’s clear that the product $\sigma \mathit{B}_{\mu\mu}$ 
depends on the particle properties (i.e., on the chosen model), 
$\mathcal{L}$ is fixed and also $\epsilon_{D}$ is fixed (depends on the detector). 
The efficiency to measure a DV depends on the distance traveled by the resonance 
before its decay, which depends on its momentum and lifetime $\tau$. If the resonance lifetime 
is very short (below $\tau \sim 0.1$ ps, which corresponds to a $c\tau$ 
lower than $\sim 30 \mu m$), 
then its decay products are named prompt, i.e. are consistent with originating from the PV 
because the detector can not distinguish such a small
distance and its discovery cannot happen thanks to a DV search. 
On the contrary, if the lifetime is bigger, then the resonance may decay at a DV. 
The lifetime $\tau$ has an 
exponential distribution which depends on the mean lifetime $\langle\tau\rangle$.
Both $\sigma$ and $\langle\tau\rangle$ depend
on the coupling constants of the BSM but, for the specific process discussed in this paper, 
they do not involve the same coupling constants (apart interference effects and higher order contributions): 
$\sigma$ depends on the quarks-gluons coupling constants at the production vertex while $\langle\tau\rangle$ 
depends on the leptons coupling constants at the decay vertex. 

There are several BSM models where a resonance with such characteristics is present and, 
regardless of the model itself, is it often possible to set independently the quarks-gluons 
coupling constants and the leptons coupling constants to have a resonance with a lifetime in 
the interesting range. The cross-section $\sigma$ could then be used to calculate the number 
of events produced in run 2 at LHC using eq. \ref{eq:numberDV}. This is the common approach in 
all LHC analyses: given a BSM model, a prediction for the number of expected events is compared with 
the SM expected number and, in case of no deviation, a limit 
is placed on the BSM parameters. The purpose of this paper is instead to 
evaluate if what has been done so
far could be improved and the key question becomes: given a 
narrow resonance which decay to a $\mu^+\mu^-$ pair with a specific lifetime 
(which depends only on the lepton coupling constants), what is the minimal cross-section 
(times the branching ratio to $\mu^+\mu^-$) $\sigma_{D}$ which may produce a discovery at the LHC run 2? 

To answer this question, it is necessary to define what is the discovery condition. 
The standard criterion for claiming a discovery in particle physics is that the observed 
effect should have the equivalent of a five standard-deviation discrepancy with the Standard Model. 
This paper will use this discovery condition but inverting eq. \ref{eq:numberDV}: 
given the mass and mean lifetime of the resonance and the SM background distribution, 
it is possible to calculate 
the expected number of SM events at the resonance mass and, using the 5$\sigma$ criterion, 
extrapolate the minimum $\sigma \mathit{B}_{\mu\mu}$ which will produce a 5$\sigma$ discrepancy 
with respect to the SM background.


\section{LLP Model and simulation}
\label{sec:llp_model}

The benchmark model used in this study is the one~\cite{Ciobanu:2005pv} implemented in 
\PYTHIA where a new $U(1)$ gauge group is considered for the production of a new neutral 
gauge boson $Z'$.
By a suitable choice of the parameters, it is possible to simulate just about any imaginable $Z'$ scenario, 
including one where the $Z'$ lifetime is big enough to produce a DV. 
To fully explore the LHC scenario, 5 different mass points have been generated from 
200 GeV to 1 TeV with 200 GeV steps. 
A proper choice of the model parameters will allow to generate a $Z’$ with a mean lifetime
in a range that produce a DV up to few meters away from the PV.
The generated events are then 
passed to the Delphes simulation where a Vertex-Fitter code~\cite{Bedeschi:2021talk} is also run 
to identify the PV and the dimuon vertex, which, depending on the model parameters, may coincide 
with the PV or generate a DV.

To estimate the LHC sensitivity to the signal, the 3 main SM background processes have been also generated 
and simulated using \PYTHIA and \MADGRAPH~\cite{Alwall:2014hca}: 
Drell-Yan production $p\bar{p} \rightarrow \gamma/Z^0 \rightarrow \mu^+\mu^-$, 
$t\bar{t}$ production and single-top production (\emph{s}-channel, \emph{t}-channel
and  $tW$). All simulated samples include an average of 35
pile-up events.
Because the target is to analyze the effectiveness
of the LHC run 2 analysis listed in table~\ref{tab:inclusiveSearch}, both the signal and the background processes
have been normalized to an integrated luminosity of $\mathcal{L}=$ 140 fb$^{-1}$.
A basic set of cuts has been used to select $pp \rightarrow \mu\mu$ events:
\begin{itemize}
  \item at least 2 opposite-charge muons with $p_{T} > 30(10)$ GeV/c for the most (second) energetic muon;
  \item $|\eta|<2.4$ for both muons;
  \item relative charged isolation defined as $\sum{p_{T}}/p_{T}^{\mu}< 0.3$ and the sum is extended to 
  all tracks with $\Delta R = \sqrt{\Delta\phi^2+\Delta\eta^2} < 0.3$.
\end{itemize}

The width of the BSM resonance depends on its lifetime and above $\langle\tau\rangle \sim 10^{-25}$ s, the $Z'$ width 
is too small to be measured with LHC detectors and the measured width is basically given by the dimuon mass resolution 
of the detector (which therefore depends only on the $Z'$ mass). The mass resolution has been extracted from the 
simulation of the $Z'$ signal and varies from 3.8 GeV for $m_{Z'} = 200$ GeV up to 54.4 GeV for $m_{Z'} = 1$ TeV.

The last ingredient which allows to calculate $\sigma_{D}$ as a function of the LLP mean lifetime $\langle\tau\rangle$
is then given by the equation
\begin{equation}
  \label{eq:tau}
  \tau = \frac{L_{xyz}}{c\beta\gamma} =\frac{mL_{xyz}}{|\vec{p}|},
\end{equation}
where $m$ is the mass of the LLP, $L_{xyz}$ is the PV-DV distance and $\vec{p}$ is the LLP momentum. 
At generator level, the mean $Z'$ lifetime $\langle\tau\rangle$ may be calculated from 
the exponential fit of the $Z'$ lifetime distribution obtained using eq.~\ref{eq:tau}. 
From an experimental point of view, equation~\ref{eq:tau} shows that, given a model which foresees a narrow resonance with 
a given mass and mean lifetime, the number of LLPs which can be detected depends on the momentum spectrum
(the larger the LLP momentum, the greater the distance traveled before the LLP decay). It's important 
to underline that the momentum spectrum of such narrow resonance produced in the $s$-channel at LHC
will be the same regardless of the chosen BSM model.

Putting together all the ingredients, the recipe to calculate $\sigma_{D}$ is the following: for each set
of the $Z'$ \PYTHIA parameters, 50000 events have been generated and simulated with Delphes. The generated
mean lifetime is calculated using the distance between the generated PV and DV 
of the $Z'$. The number of signal events $N_{S}=\sigma_{D}\epsilon\mathcal{L}$
which pass a specific selection for $\mathcal{L}=$140 fb$^{-1}$, is obtained
from the integral of the invariant dimuon mass distribution over the range $[m-2\Gamma,m+2\Gamma]$
where $\Gamma$ is the experimental width of the $Z'$. 
The number of background events $N_{B}$ is obtained from the background only distribution 
integrated over the same mass range used for the signal sample. 
Finally, the discovery cross-section $\sigma_{D}$ 
is calculated from
\begin{equation}
  \label{eq.xsec_discovery}
  \sigma_{D} = \frac{5\sqrt{N_{B}}}{\epsilon_D \mathcal{L}},
\end{equation}
where the factor 5 is for the common 5$\sigma$ discovery condition.

\section{Results}
Table~\Ref{tab:inclusiveSearch} indicates clearly that two quantities are mainly used
to identify a DV: the muons transverse impact parameter $d_0$ for both muons and
the  distance $L_{xy}$. 
Different selection cuts are applied to these 2 quantities or to their significance. A full scan
of these parameters together with the 3D distance $L_{xyz}$ and its significance could 
give an indication on what is the best approach for a LLP search. For each variable, 
the discovery cross-section is calculated using the base cuts described in section~\ref{sec:llp_model}
together with a single cut on one of the 6 variables considered. 
The chosen ranges for each variable are the following: $d_0 \in [0,0.1]$ mm, $L_{xy}$ and $L_{xyz}$ in
$[0,2]$ mm and all significance $|d_0/\sigma_{d_0}|$, $L_{xy}/\sigma_{L_{xy}}$
and $L_{xyz}/\sigma_{L_{xyz}}$ in $[0,20]$.

\begin{figure}[hbtp]
  \begin{minipage}{0.45\textwidth}
  \centering
  \includegraphics[width=0.98\textwidth]{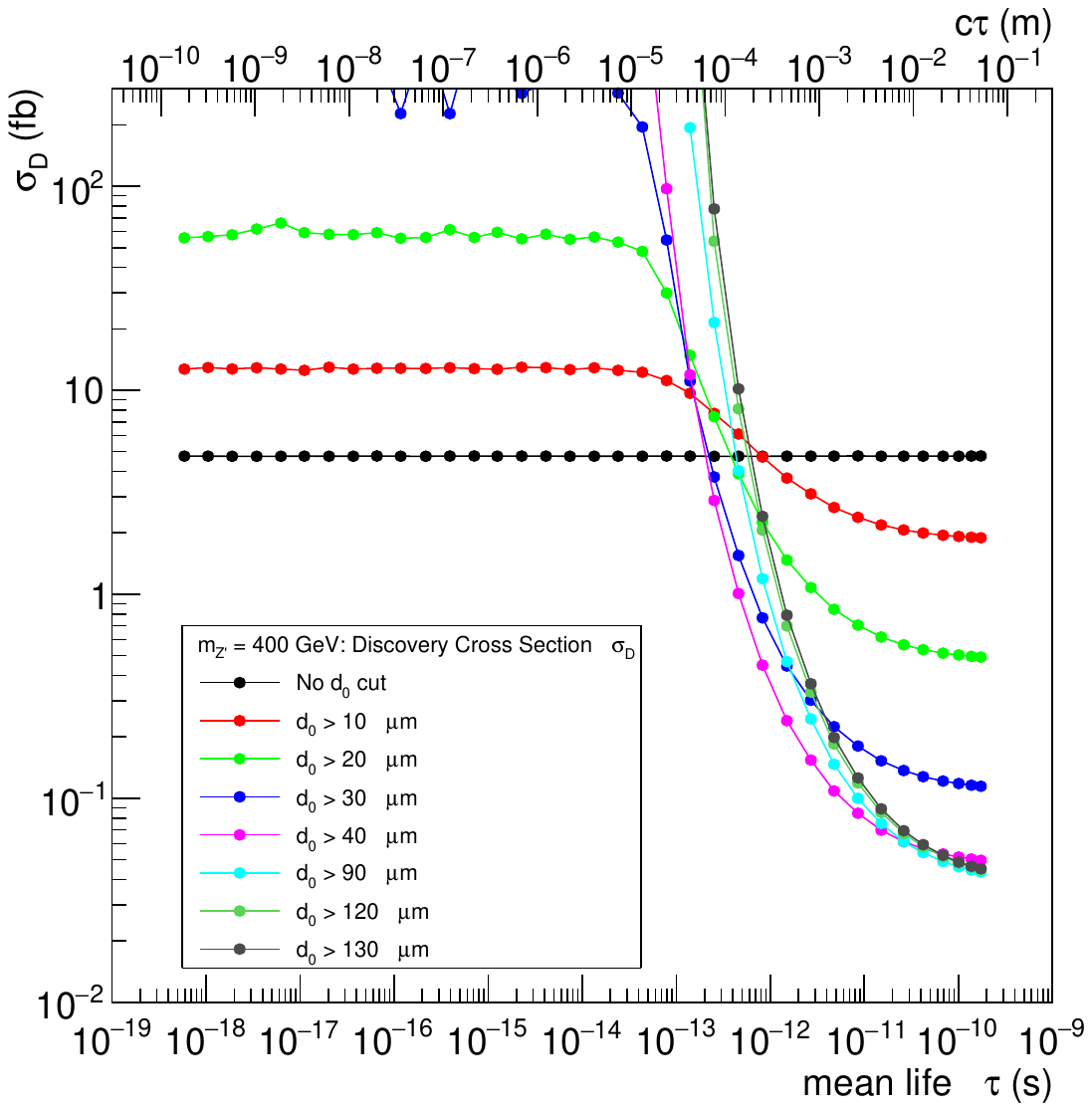}
  \caption{Discovery cross-section $\sigma_{D}$ for different transverse impact parameter cuts for a 400 GeV $Z'$.}
  \label{fig:d0_scan_400}
  \end{minipage}
\end{figure}

As an example, fig.~\ref{fig:d0_scan_400} show the discovery cross-section $\sigma_{D}$ for a 400 GeV $Z'$ as a 
function of its mean lifetime and transverse impact parameter cut used in the selection. 
The area above each curve represents the combination of $Z'$ parameters that may lead to a discovery at LHC while 
the area below indicates a combination of $Z'$ parameters which gives too few events to allow for a discovery using
the 5$\sigma$ criterion.
It's evident that below $\langle\tau\rangle \backsimeq 0.1$ ps (which corresponds to a $c\langle\tau\rangle \backsimeq 30 \mu$m) 
the distance between the PV and the DV is too small for the actual generation of LHC detectors and the 2 muons from 
the $Z'$ decay are prompt: in this case the best approach is to avoid any cut on the transverse impact parameter.
Above $\langle\tau\rangle \backsimeq 30$ ps (which corresponds to a $c\langle\tau\rangle \simeq 10$ mm), the $\sigma_{D}$ 
shape is flat and there is no improvement in going to a $d_0$ cut bigger than $\simeq 90\ \mu$m.
A full scan of the 6 variables listed above allows to identify the best variable/cut value pair for different
mean lifetime to maximize the discovery potential of a $Z'$ resonance with a mass up to 1 TeV.

\begin{figure}[hbtp]
  \begin{minipage}{0.45\textwidth}
  \centering
  \includegraphics[width=0.98\textwidth]{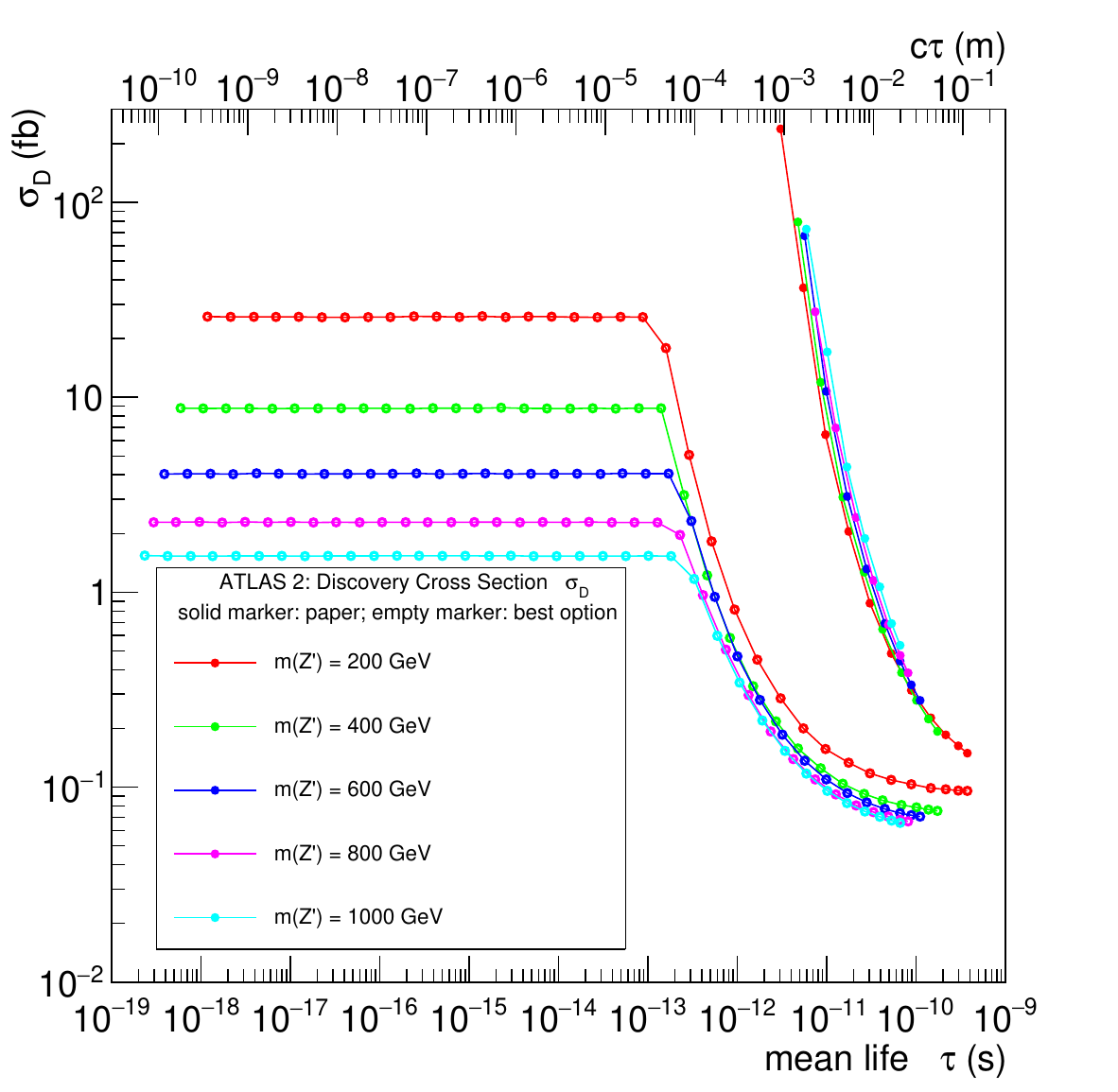}
  \caption{Discovery cross-section $\sigma_{D}$ for the best combination and "ATLAS 2" paper~\cite{ATLAS:2019fwx}.}
  \label{fig:compL1}
  \end{minipage}
\end{figure}

A better understanding of the discovery potential could be seen in fig.~\ref{fig:compL1} where (as an example) the
best combination obtained from the full 6-variables scan (extended to the 5 $Z'$ masses discussed here) is compared
to the $\sigma_{D}$ obtained using the set of cuts applied in one of the ATLAS papers 
normalized to $\mathcal{L}=$140 fb$^{-1}$
(see table~\ref{tab:inclusiveSearch}). 
This comparison shows that the set of cuts used in ref.~\cite{ATLAS:2019fwx} is not optimized for 
a $Z'$ with a $c\langle\tau\rangle$ below few cm. 

\section{The $\cos\phi_{3D}$ approach}

In the previous section it was shown that LHC experiments can improve LLP searches by optimizing the main variables cuts 
used to select signal events.
The most interesting region to study is where the mean path of the LLP resonance is at most few mm. 
In this region, however, the SM background due to detectors effects is not negligible, and therefore it is necessary to develop 
an appropriate strategy to evaluate it. All the previous plots have been produced assuming that simulated events represents correctly
data distributions but for the 6 variables considered, it is known that
also full simulated events are not good enough to represent correctly background distributions and all the papers listed 
in Tab.~\ref{tab:inclusiveSearch} prefer to evaluate it directly from data. It is evident that
the final result will depend on the chosen procedure for the background evaluation and 
the capability to keep under control all the systematic uncertainties.

\begin{figure}[hbtp]
  \begin{minipage}{0.45\textwidth}  
  \centering
  \includegraphics[width=0.98\textwidth]{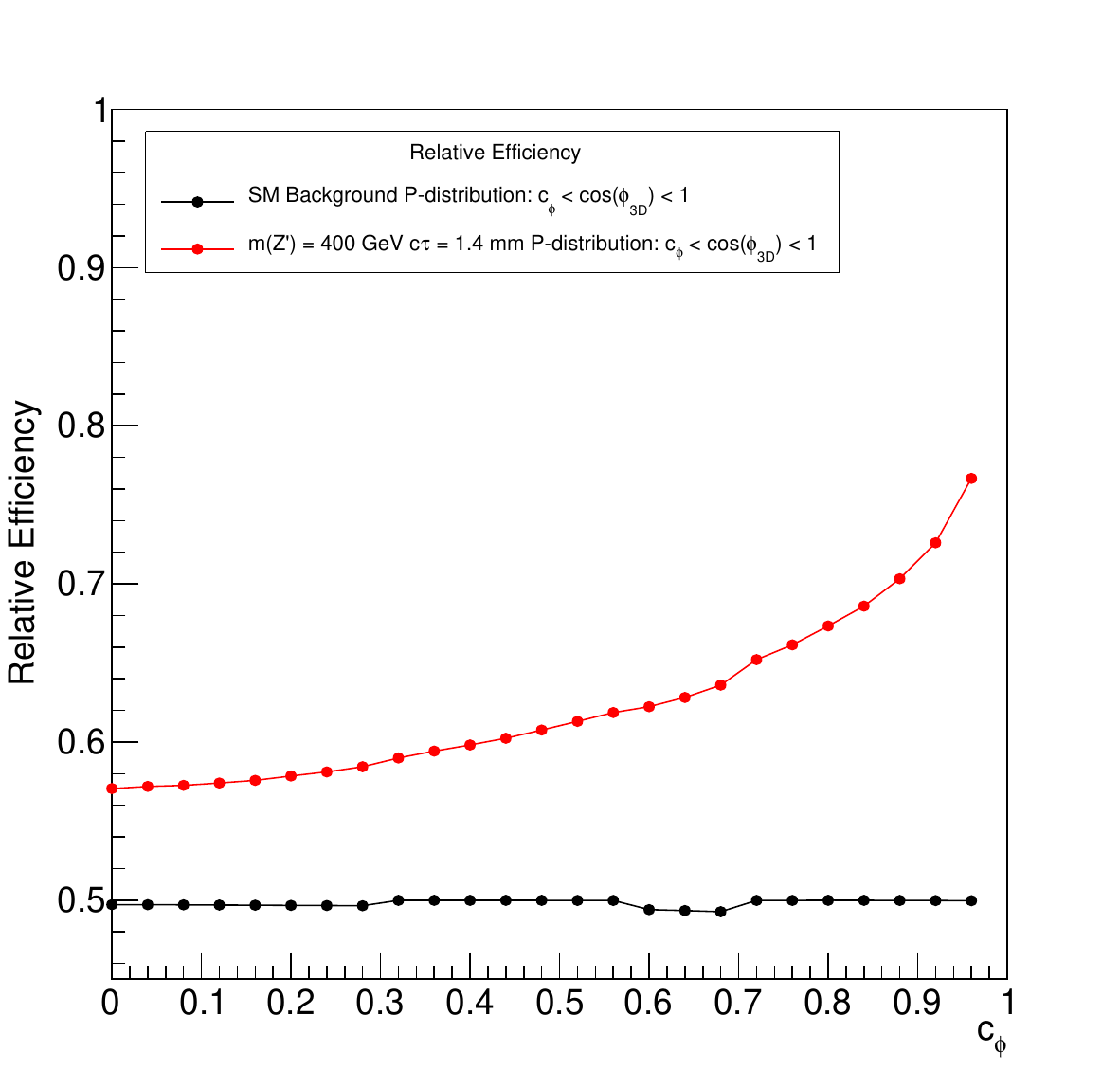}
  \caption{Relative efficiency for the P- and M-distribution as a function of $c_{\phi}$.}
  \label{fig:plot103_RelEff}
  \end{minipage}
\end{figure}

The new proposed approach will use the $\phi_{3D}$ angle introduced in section \ref{sec:intro}
for reducing at a minimum the systematic uncertainties: above $m(\mu\mu) = $ 200 GeV there are no SM resonances
and the $\vec{\Delta}_{3D}/\vec{p}_{\mu\mu}$ vector directions will be fully uncorrelated.
The distribution of $\cos\phi_{3D}$ would be fully symmetric around $\cos\phi_{3D}=0$ while
for a LLP decaying far from the PV, the $\cos\phi_{3D}$ distribution will be strongly peaked at 1.
The dimuon invariant mass distribution could be therefore divided in 2 different distributions:
the Plus-distribution (P-distribution: dimuon invariant mass where $ c_{\phi} \leq \cos\phi_{3D} \leq 1$) and the Minus-distribution 
(M-distribution: dimuon invariant mass where $ -1 \leq \cos\phi_{3D} \leq -c_{\phi}$), where $c_{\phi}$ may be chosen in the range $[0,1]$.
For SM background events, the P- and M-distribution will be identical within statistical errors, while a LLP will produce a strong
peak in the P-distribution only. Fig.~\ref{fig:plot103_RelEff} shows this effect for SM background and 
a 400 GeV $Z'$ with $c\langle\tau\rangle = 1.4$ mm: for SM background, the relative efficiency is always 50\% (within statistical error)
and the P/M-distribution are identical; for a LLP, the higher the cut on the $c_{\phi}$, the greater will be
the P- relative efficiency with respect to the M-distribution (as an example, using Fig.~\ref{fig:plot103_RelEff}, for $c_{\phi}$ = 0.90, 
the relative efficiency will be 72\% for the P- and 28\% for the M-distribution).

\begin{figure}[hbtp]
  \begin{minipage}{0.45\textwidth}  
  \centering
  \includegraphics[width=0.98\textwidth]{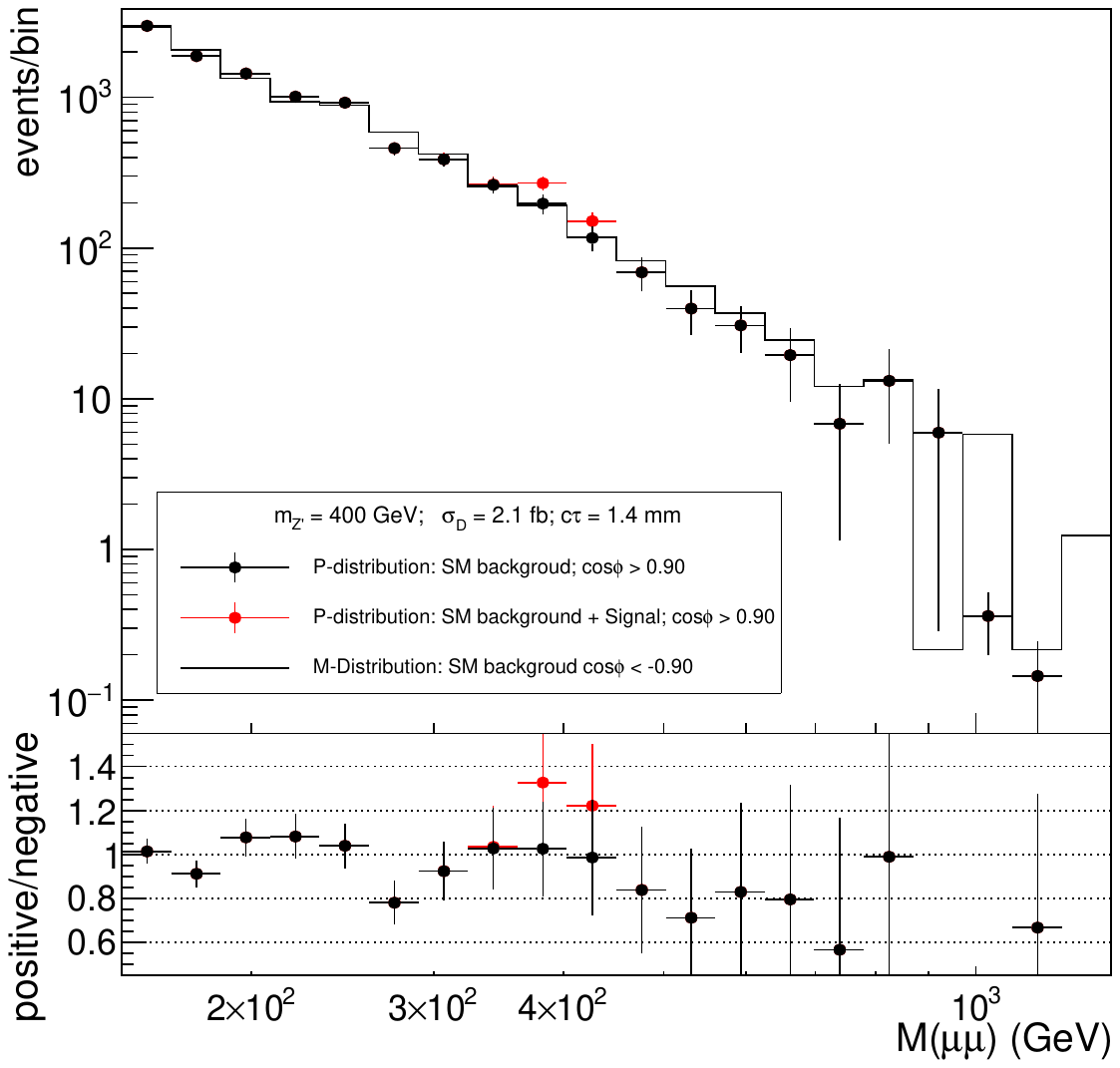}
  \caption{Dimuon invariant mass P-distributions for SM background events only and with a 400 GeV $Z'$ with $c\langle\tau\rangle = 1.4$
     mm superimposed ($|d_0/\sigma_{d_0}| \ge 1.0$ and $c_{\phi_{3D}} = 0.90$). The M-Distribution with or without the $Z'$ signal 
     are almost identical and cannot be distinguished in the plot}
  \label{fig:pm_distribution}
  \end{minipage}
\end{figure}

An example is shown in fig.~\ref{fig:pm_distribution} where a 400 GeV $Z'$ 
signal with $c\tau = 1.4$ mm and $\sigma B_{\mu\mu} = 2.1$ fb has been added to the SM background. The selection applied 
is $|d_0/\sigma_{d_0}| \ge 1.0$ and $ c_{\phi_{3D}} = 0.90$.
This approach offers an undisputed advantage
with respect to other methods: almost all systematic uncertainties could be neglected because they will affect at the same level
both the P- and M-distributions. The only check needed (and eventually considered as a systematic effect) is to verify
that the P- and M-distributions are identical within statistical errors.

The discovery cross-section can be obtained using 
again eq. \ref{eq.xsec_discovery} calculating the number of events $N_P$ and $N_M$ over the range $[m-2\Gamma,m+2\Gamma]$
in the P- and M-distributions:
$N_{P/M} = B_{P/M} + \sigma_{D}\mathcal{L}\epsilon_{P/M}$ where $B_{P/M}$ are respectively the SM background P- and M-distribution, 
$\sigma_{D}\mathcal{L}\epsilon_{P/M}$ are the $Z'$ P- and M- contribution 
and the significance is $(N_P - N_M)/\sqrt{N_M}$. Requesting
the $5\sigma$ discovery condition will allow calculating $\sigma_{D}$ as a function of the $Z'$ mass and lifetime. Six different
$c_{\phi_{3D}}$ values in the range $[0,0.98]$ have been used together with the 6 variables already discussed. A cross-check
of the goodness of the P/M method can be done using a profile likelihood ratio as a statistical test to calculate the significance
of the signal. Assuming a Gaussian signal with mean value equal to 400 GeV and width equal to the mass resolution 
(11.4 GeV for a 400 GeV $Z'$), the background given by the M-distribution and the data by the P-distribution, the 
calculated significance is 5.12 in agreement with the $5\sigma$ discovery condition.

\begin{figure}[hbtp]
  \begin{minipage}{0.45\textwidth}  
  \centering
  \includegraphics[width=0.98\textwidth]{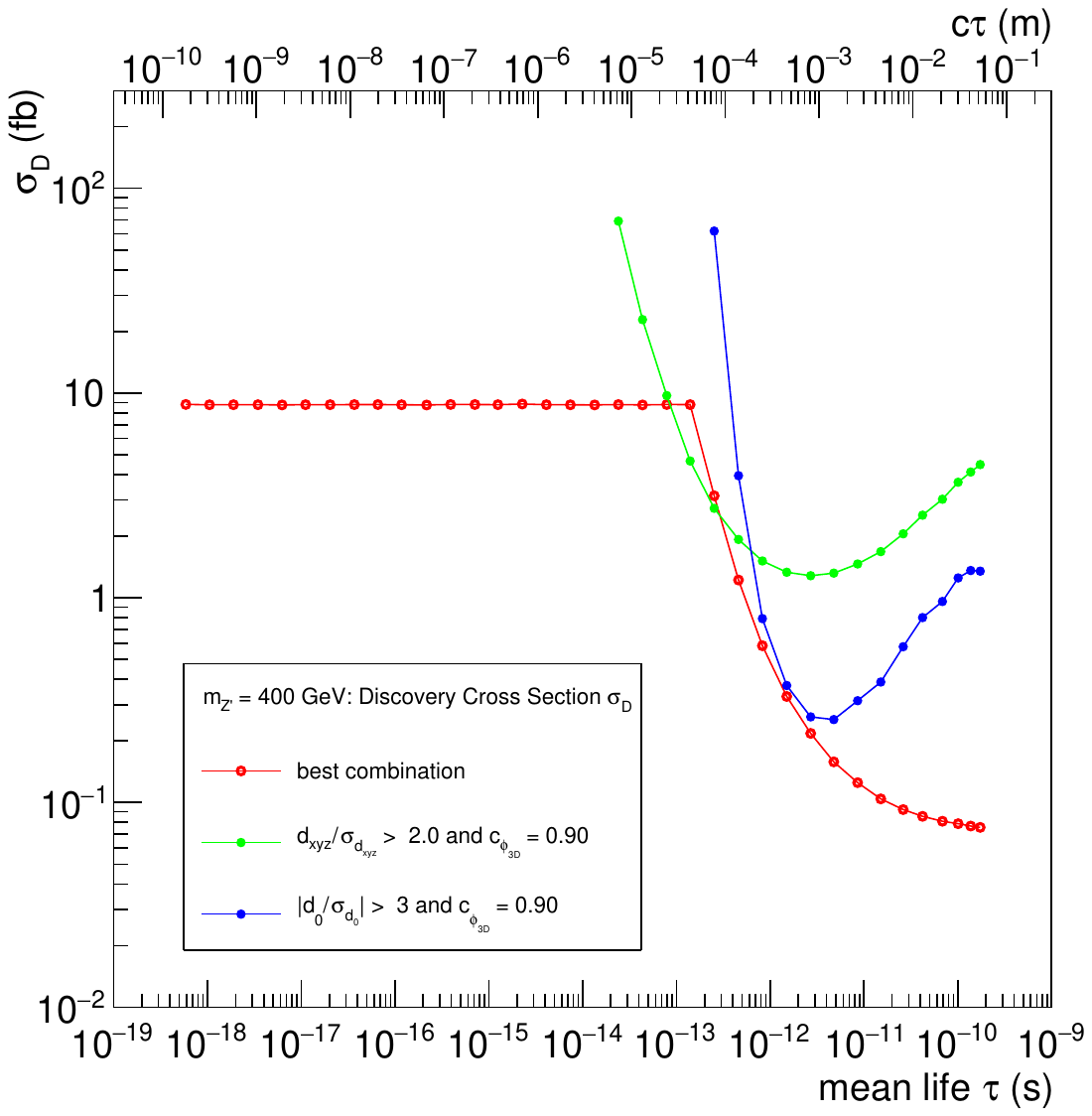}
  \caption{Discovery cross-section $\sigma_{D}$ for a 400 GeV $Z'$ resonance: ideal best combination and 2 possibile
    choice of the $\cos\phi_{3D}$ approach.}
  \label{fig:bestAngle}
  \end{minipage}
\end{figure}

Fig.~\ref{fig:bestAngle}
compares the best combination and 3 different selections using the P/M approach
for a 400 GeV $Z'$: 
again, below $\langle\tau\rangle \simeq 0.1$ ps, there is no improvement in using any specific selection: all muons are prompt,
and DVs are not distinguishable from the PV. Above $\simeq 10$ ps, there are virtually no SM background events: all reconstructed DVs
are due to detector effects or mis-reconstructed cosmic rays. The full scan of the 6 considered variables shows that
the significances $|d_0/\sigma_{d_0}|$ and $L_{xyz}/\sigma_{L_{xyz}}$ are the most powerful quantities to selects LLPs.
In the region between 0.1 and 10 ps however, the P/M approach has many advantages with respect to previous published methods: $\sigma_{D}$ is
comparable with the ideal best combination but virtually all systematic uncertainties 
normally studied could be neglected because affect identically both the P- and M-distribution.

\section{Conclusions}
Searches of LLPs are widely used at LHC experiments for BSM particles discovery. 
This paper has shown that a careful study is needed to fully exploit tracker detector potentialities for a 
DV within few mm from the PV and
a full-simulation study is needed to confirm the outcome of this paper:
the best approach to search for neutral resonances which decay within a few mm from the PV
is to use $|d_0/\sigma_{d_0}|$ and $L_{xyz}/\sigma_{L_{xyz}}$ for pre-selecting the muon pairs.
However, also full simulation for these quantities are far from being perfect and a robust search will
have to be based on data only. Finally, a new approach has been proposed which is based on exploiting the properties of the
$\phi_{3D}$ angle. This new approach, fully based on data, allows reducing at a minimum the systematic uncertainties 
while keeping at best the capability to detect a BSM resonance which decays nearby the PV.

\section*{Acknowledgments} 
I'm grateful to
Gian Mario Bilei,
Livio Fan\'o,
Michele Gallinaro,
Valentina Mariani,
Simone Pacetti,
Matteo Presilla and
Alessandro Rossi 
for many useful discussions and comments on the manuscript.

\bibliography{plb05}{}
\bibliographystyle{elsarticle-num}

\end{document}